\documentclass[a4paper,fleqn]{cas-dc}

\usepackage[numbers]{natbib}

\usepackage{algorithm}
\usepackage{algpseudocode}
\usepackage{pgfplots}
\usepackage{footmisc}
\usepackage{epsfig}
\usepackage{float}
\usepackage{xcolor}
\algdef{SE}[SUBALG]{Indent}{EndIndent}{}{\algorithmicend\ }%
\algtext*{Indent}
\algtext*{EndIndent}

\def\tsc#1{\csdef{#1}{\textsc{\lowercase{#1}}\xspace}}
\tsc{WGM}
\tsc{QE}
\tsc{EP}
\tsc{PMS}
\tsc{BEC}
\tsc{DE}

\begin{document}
\let\WriteBookmarks\relax
\def\floatpagepagefraction{1}
\def\textpagefraction{.001}
\shorttitle{Multi-stage MPC for slug flow crystallizers using uncertainty-aware surrogate models}

\shortauthors{C.R. Johnson et~al.}

\title [mode = title]{Multi-stage model predictive control for slug flow crystallizers using uncertainty-aware surrogate models}                      



%
\author[1]{Collin R. Johnson} 
\cormark[1]
\author[1]{Stijn de Vries}
\author[2]{Kerstin Wohlgemuth} 
\author[1]{Sergio Lucia} 






\affiliation[1]{organization={Chair of Process Automation Systems, TU Dortmund University},
            addressline={Emil-Figge-Str. 70}, 
            city={Dortmund},
            postcode={44227}, 
            country={Germany}}
\affiliation[2]{organization={Laboratory of Plant and Process Design, TU Dortmund University},
            addressline={Emil-Figge-Str. 70}, 
            city={Dortmund},
            postcode={44227}, 
            country={Germany}}





\nonumnote{This work was funded by the Deutsche Forschungsgemeinschaft (DFG, German Research Foundation) – 504676854 – within the Priority Program “SPP 2364: Autonomous processes in particle technology”.}

\begin{abstract}
This paper presents a novel dynamic model for slug flow crystallizers that addresses the challenges of spatial distribution without backmixing or diffusion, potentially enabling advanced model-based control. The developed model can accurately describe the main characteristics of slug flow crystallizers, including slug-to-slug variability but leads to a high computational complexity due to the consideration of partial differential equations and population balance equations. For that reason, the model cannot be directly used for process optimization and control. To solve this challenge, we propose two different approaches, conformalized quantile regression and Bayesian last layer neural networks, to develop surrogate models with uncertainty quantification capabilities. These surrogates output a prediction of the system states together with an uncertainty of these predictions to account for process variability and model uncertainty. We use the uncertainty of the predictions to formulate a robust model predictive control approach, enabling robust real-time advanced control of a slug flow crystallizer. 
\end{abstract}



\begin{keywords}
Model predictive control \sep Uncertainty quantification \sep Surrogate modeling \sep Continuous crystallization
\end{keywords}

\maketitle

\section{Introduction}\label{section:Introduction}

Autonomous process control offers many advantages for the process industry \citep{schwenzerReviewModelPredictive2021} since processes can be run both safer and more efficiently \citep{zengSafetyCriticalModelPredictive2021}, operating closer to physical limits established by product specifications. This can lead to a reduction in costs, energy, and resource consumption. Autonomous processes are usually realized by model-based control approaches, such as model predictive control (MPC), which is based on forecasting the behavior of the process using a model. Obtaining a detailed model in the first place is thus one of the main challenges to achieve autonomous processes \citep{schwenzerReviewModelPredictive2021}.

An area where autonomous process operation can lead to considerable improvements is crystallization \citep{yazdanpanahHandbookContinuousCrystallization2020}. Continuous crystallization promises more reliable and efficient processes, at the cost of more difficult modeling and automation. A promising apparatus in the field of continuous crystallization is the slug flow crystallizer \citep{termuhlenContinuousSlugFlow2021, jiangContinuousFlowTubularCrystallization2014}, which can have important advantages with respect to batch crystallization especially for components that require low production rates, such as active pharmaceutical ingredients. In slug flow crystallization, backmixing can be avoided by the right choice of tubing material and segmentation medium for the liquid phase, enabling single slugs which do not mix \citep{kufnerStrategyFastDecision2022}. By avoiding backmixing, the slug flow crystallizer offers some advantages on the process side, i.e. reducing axial dispersion and improving particle suspension \citep{termuhlenContinuousSlugFlow2021}. Because the slug flow crystallizer is a spatially distributed system without backmixing or diffusion, it is difficult to model. The dependence on time and the additional description of the particle size distribution lead to a partial differential equation as a function of three variables (time, space, and particle size). 
Until now, models for the slug flow crystallizer have circumvented these problems by modeling a single slug as a batch crystallizer traveling through the slug flow crystallizer, leading to steady-state models \citep{kufnerModelingContinuousSlug2023, rascheMathematicalModelingOptimal2016b}. In this work, we present a fully dynamic model for the slug flow crystallizer, which by design exhibits no diffusion. 

The price to pay for the accurate dynamic description of the slug flow crystallizer is a high computational cost. As a result, the proposed model cannot be directly used for online optimization and control. We propose to use the detailed model to generate data to train a surrogate model that accurately approximates the detailed model.
Data-based surrogate models have been widely studied in recent years. 
A popular framework for data-based modeling is the sparse identification of nonlinear dynamic systems methodology (SINDy)\citep{bruntonDiscoveringGoverningEquations2016}. Here, a feature library is selected in advance and only the most important features that can best explain the data are used. The resulting model can potentially be interpretable, but generating the feature library containing the important features is difficult in general. Instead, in this work we use neural networks that determine the feature space during training based on data. The resulting models are less interpretable, but prior knowledge of the feature space is not necessary for model training. 

A decisive disadvantage of surrogate models is the loss of fundamental validity. The predictions of the underlying first-principle model are valid at least until a model assumption is violated. The approximation, on the other hand, is only valid for interpolation, which is an abstract concept in a high-dimensional space \citep{balestrieroLearningHighDimension2021}. When these models are integrated into an optimization-based controller, the optimization solver can exploit regions of the surrogate models that are mathematically advantageous but physically meaningless in reality. It is therefore very important that surrogate models, especially when black-box neural networks are used, can quantify the uncertainty of their predictions.
Gaussian processes are a typical approach used to quantify uncertainty of predictions for data-based models \citep{rasmussenGaussianProcessesMachine2006}. However, they do not scale well for large amounts of data, which is typically required for very nonlinear spatially distributed systems, as it is the case in slug flow crystallization. The main approach based on neural networks that can quantify the uncertainty of its predictions is Bayesian neural networks (BNNs) \citep{jospinHandsonBayesianNeural2020}. BNNs are similar to standard neural networks, but the trainable weights of the linear transformations are assumed to be Gaussian distributed, leading to Gaussian distributed predictions. Unfortunately, this approach leads to computationally intractable problems. A compromise between BNNs and standard neural networks can be achieved by Bayesian last layer neural networks (BLLs) \citep{fiedlerImprovedUncertaintyQuantification2023a}. Here, only the weights of the last layer are Gaussian distributed. By choosing a linear activation function for the last layer, which is common practice in regression, the problem becomes computationally tractable. Also very recently, conformalized quantile regression (CQR) \citep{romanoConformalizedQuantileRegression2019} has emerged as a powerful approach to quantify uncertainty of predictions in the area of machine learning. In CQR quantiles are trained to capture the shape of the variability and then a conformalization step is performed on unseen data to estimate the approximation error.

This work has two central contributions. First, we develop for the first time a full dynamic model for the slug flow crystallizer, which can consider important characteristics such as slug-to-slug variability and non-constant slug velocities over the length of the crystallizer. The presented model produces solutions free of numerical diffusion, which is usually difficult to achieve with standard discretization techniques but is especially important for the slug flow crystallizer due to the absence of backmixing and therefore the absence of physical diffusion. Second, we propose the use of surrogate models with neural networks, expanded by the integration of conformalized quantile regression and Bayesian last layers to quantify the uncertainty of the predictions. Uncertainty quantification of surrogate models is crucial to ensure that approximation errors of the surrogate, parametric uncertainties or inherent process variability can be explicitly taken into account when designing a model-based controller. To exploit the uncertainty quantification we use a robust nonlinear model predictive control approach based on scenario trees, leading to a real-time capable, efficient, and robust control of the slug flow crystallizer. The work is an extension of the work presented in  \cite{johnsonRobustNonlinearModel2024}, where BLL models were used to consider the approximation error in the controller for a simple crystallization system. We extend the investigation to the significantly more complex slug flow crystallizer system for which we develop a new dynamic model. Process variability is particularly important for the slug flow crystallizer. Since each slug contains a population of particles, the particle size distribution varies significantly from slug to slug. In addition, the variability itself varies significantly for changing process parameters. We directly consider process variability in our control scheme and adaptively operate closer or farther from constraints by using CQR models in our controller. In addition, we use BLL models in our controller to operate efficiently given model uncertainty due to lack of data.

This work is structured as follows: In Section~\ref{section:Model}, we introduce a new dynamic model for the slug flow crystallizer. In Section~\ref{section:SurrogateModels}, we train neural network-based surrogate models to obtain optimization-friendly models. In Sections~\ref{section:MPC} and \ref{section:Results}, we show how the models and their uncertainty quantification are used in a robust model predictive control framework and we evaluate the results with thorough simulation studies. The paper is concluded in Section~\ref{section:Conclusion}.

\section{Model development for the slug flow crystallizer}\label{section:Model}

The slug flow crystallizer is a challenging system for modeling, since it is distributed in the spatial length and the particle size distribution. Additionally, the lack of backmixing or diffusion between the liquid slugs is difficult for standard discretization schemes, since these usually exhibit at least some degree of numerical diffusion. In addition, the flow in the slugs flow crystallizer exhibits an increasing velocity profile over the length of the crystallizer. In the crystallizer, wall friction, liquid viscous resistance, as well as interfacial tension, lead to a pressure drop. The air slugs inside the crystallizer are not incompressible and, for that reason, they expand. Subsequently, the expansion of the gas slugs leads to an increasing velocity over the length of the crystallizer. The increasing velocity profile influences the residence time and therefore also the crystallization process that occurs in the liquid slugs.

Although there are only few works on modeling the slug flow crystallizer, the common approach is to treat individual slugs as batch crystallizers traveling through the crystallizer, solving the issue of describing a complete lack of backmixing and diffusion. In \cite{rascheMathematicalModelingOptimal2016b} and \cite{mozdzierzMathematicalModelingExperimental2021b} individual slugs are modeled as batch crystallizers. The slug flow crystallizer is tempered with different tempering baths, either with constant temperature or countercurrent flow. Using knowledge of the residence time in the baths, the differential equations describing the single slugs can be solved, leading to the evolution of the states as function of the residence time or the length of the crystallizer. A different approach can be seen in \cite{kufnerModelingContinuousSlug2023}. The slug flow crystallizer here consists of a co-current tube-in-tube concept as shown in Figure~\ref{figure:SystemSketch}. A steady-state model is obtained by again treating single slugs as batch crystallizers and subsequently transforming the time derivative into a spatial derivative using the velocity. Using the velocity for the transformation to spatial derivatives enables the consideration of the velocity profile. The disadvantage of this approach is that the resulting model is static, and therefore not suitable for model-based control, which requires a dynamic model.

\subsection{Novel dynamic modeling approach}

The presented approach for modeling the slug flow crystallizer uses different techniques to consider gas expansion and velocity profile while still resulting in a dynamic model. The slug flow crystallizer under consideration for the proposed model is described in \cite{termuhlenContinuousSlugFlow2021, termuhlenCharacterizationSlugFormation2019}. The dynamic model developed in this work uses correlations from the static model in \cite{kufnerModelingContinuousSlug2023} which describe hydrodynamic and crystallization phenomena of the slug flow crystallizer.
\begin{figure}
    \centering
    \includegraphics{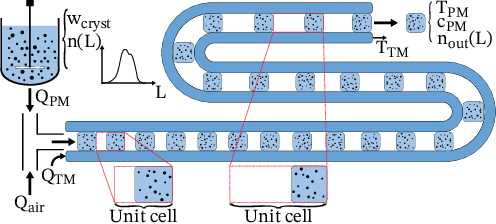}
    \caption{Sketch of the slug flow crystallizer system under consideration.}
    \label{figure:SystemSketch}
\end{figure}
A sketch of the system can be seen in Figure~\ref{figure:SystemSketch}. The slug flow crystallizer under consideration consists of two different tubes, where the outer tube is fed with the tempering medium. The inner tube is fed with the process medium and air that forms the slug flow. Accordingly, for the presented model, we consider different modeling strategies for the outer tube containing the tempering medium and for the inner tube containing the segmented process medium. The volume flows of the tempering medium, the process medium and the air are given by $Q_{TM}$, $Q_{PM}$, and $Q_{\text{air}}$. The process medium contains seed particles according to the weight fraction $w_{\text{cryst}}$ and the particle size distribution $n(L)$. We assume concentration and temperature to be constant at the inlet. At the outlet, we consider the temperatures of the tempering medium $T_{TM}$, and of the process medium $T_{PM}$ as well as the concentration of the process medium $c_{PM}$ and the particle size distribution $n_{\text{out}}(L)$.
The partial differential equation for the temperature of the outer tube is given by the one-dimensional convection-diffusion equation with source terms for heat transfer:
\begin{align}\label{eq:PDETemperingMedium}
    \frac{\partial T_{TM}}{\partial t} &= v\frac{\partial T_{TM}}{\partial z}+D\frac{\partial^2 T_{TM}}{\partial z^2}+s(T_{TM}),
\end{align}
where $T_{TM}=f(t,z)$ is the temperature of the outer tube and $z$ is the coordinate of the spatial length. The constant velocity is given by $v$ and the diffusion coefficient is given by $D$. The source term $s(T_{TM})$ contains heat transfer between the inner tube and the outer tube, as well as between the outer tube and the environment. \eqref{eq:PDETemperingMedium} is solved using the finite volume scheme with the $5$-th order weighted essentially non-oscillatory method (WENO) \cite{liuWeightedEssentiallyNonoscillatory1994} to compute the convective flow and central differences for the diffusion contribution.

For the process medium, the convective flow is solved by considering unit cells as introduced in \cite{kufnerModelingContinuousSlug2023}. The concept of a unit cell is shown in Figure~\ref{figure:SystemSketch}. A unit cell is comprised of a gas slug and a liquid slug. The model considers individual unit cells traveling through the slug flow crystallizer. The liquid slug in each unit cell is modeled as single batch crystallizers. However, these simulated individual batches do not necessarily coincide with the actual physical slugs in the crystallizer. A solution scheme similar to the sequencing method \cite{renouSolutionConvectionDispersion2003} is used. In the sequencing method, the simulation time step, the flow velocity, and the width of the finite volumes are coupled, leading to diffusion-free numerical solutions. Subsequently, convection, diffusion, and reaction (corresponding to crystallization and heat transfer for the slug flow crystallizer) are solved sequentially, which leads to accurate results assuming sufficiently small time steps. In Figure~\ref{figure:SystemSketch} the comparison of unit cells at the inlet and at the end of the crystallizer is shown. While liquid slugs are considered incompressible, gas slugs grow due to gas expansion, leading to an increasing velocity profile over the length of the crystallizer. Because the velocity of the slug flow crystallizer varies over time and over the length of the crystallizer, it is necessary to adapt the standard sequencing method. Hence, for the process medium, no fixed discretization scheme is used. Instead, at each time step a new individual batch is introduced at the inlet of the slug flow crystallizer. All batches already present in the slug flow crystallizer are then advanced according to the local velocity and time step. The gas expansion of the gaseous slugs along the length of the crystallizer is computed using the ideal gas law. The pressure drop is assumed to be linear along the length, and the pressure drop is calculated using the correlation from \cite{kufnerModelingContinuousSlug2023} which considers the single-phase as well as multi-phase pressure drop. An initial velocity at the inlet of the crystallizer is computed based on the volume flows of process medium and air. Subsequently, the velocity profile is computed using the degree of expansion of the gas slugs over the length of the crystallizer.

\begin{figure}
    \centering
    \includegraphics{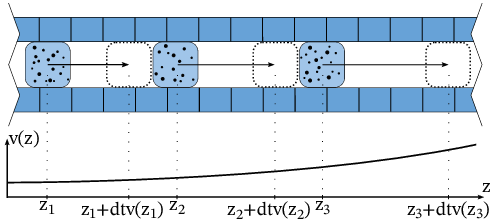}
    \caption{Illustration of the new dynamic slug flow crystallizer model. The outer tempering medium is discretized into static finite volumes. The inner process medium is modeled using batch crystallizers which are advanced through the crystallizer according to their local velocity $v(z_i)$. Depicted are three different slugs. The dotted slugs represent the respective position at the next time step.}
    \label{figure:ModelSketch}
\end{figure}

Algorithm~\ref{alg:cap} summarizes how the convective flow in the process medium is solved.
\begin{algorithm}
\caption{Solution for the process medium.}\label{alg:cap}
\begin{algorithmic}
\State $\text{slugs}=[\text{slug}_0, \dots, \text{slug}_N]$
\State $\text{slug}_i = \{z_i, \, m_i, \, c_i, \, T_i, \, n_{\text{pop},i}\}$ \\
\Require 
\State $n_{\text{steps}}$, $dt$, $\dot m_{PM}$, $Q_{\text{air}}$, $p_{\text{in}}$, $L_{\text{SFC}}$, $\text{slugs}$
    \State $\text{step}_i \gets 0$
    \While{$\text{step}_i \leq n_{\text{steps}}$}
        \State $\text{get} \, v=f(z)$ \Comment{Calculate velocity profile}
        \For{$\text{slug}_i \, \textbf{in} \, \text{slugs}$} \Comment{Go through all slugs}
            \State $z_i \gets z_i + dt \, v(z_i)$ \Comment{Advance slug}
            \If{$z_i > L_{\text{SFC}}$}
                \State $\text{delete slug}_i$
                \State break
            \EndIf
            \State $\text{update} \, c_i, \, T_i, \, A_i, \, n_{\text{pop},i}$ \Comment{Solve ODEs}
        \EndFor
        \State $\textbf{initialize new slug}$
        \Indent
            \State $z_{\text{new}} \gets 0$
            \State $m_{\text{new}} \gets dt \, \dot m_{PM}$
            \State $c_{\text{new}} \gets c_{\text{in}}$
            \State $T_{\text{new}} \gets T_{\text{in}}$
            \State $n_{\text{pop,new}}$ \Comment{Sample from init. distribution}
        \EndIndent
        \State $\textbf{end initialize new slug}$
        \State $\text{slugs} \gets \text{slug}_{\text{new}} + \text{slugs}$ \Comment{Add new slug at inlet}
        \State $\text{step}_i \gets \text{step}_i+1$
    \EndWhile
\end{algorithmic}
\end{algorithm}
The process medium is modeled as a list of individual slugs, each containing values for position and states. The position of slug $i$ is denoted by $z_i$. The states are the mass $m_i$, the concentration $c_i$, the temperature $T_i$, and the Monte Carlo particle size distribution $n_{\text{pop},i}$. The slug is first advanced according to the local velocity which corresponds to an explicit Euler solution of the convective flow as seen in Figure~\ref{figure:ModelSketch}.
The slugs are checked to determine if their position is still within the crystallizer, and the states of each respective slug are advanced in time. The differential equations for process medium temperature $T_{PM,i}$, concentration $c_i$, and particle size population $n_i$ for the $i$-th slug are given by:
\begin{align}
    \frac{d T_{PM,i}}{dt} &= \frac{U_{PM,TM}A_i(T_{TM}-T_{PM,i})}{m_ic_P}, \label{eq:TempPM} \\
    \frac{dc_i}{dt} &= -\frac{3\rho_{\text{cryst}}k_vG\mu_2}{m_i}, \label{eq:ConcPM} \\
    \frac{\partial n_i}{\partial t} &+ G\frac{\partial n_i}{\partial L} = B_{\text{Agg}}-D_{\text{Agg}}, \label{eq:PBE}
\end{align}
where $m_i$ and $A_i$ correspond to the mass of the slug and the area available for the heat transfer of the slug. The heat transfer coefficient from process medium to tempering medium is given by $U_{PM,TM}$ and the specific heat capacity is $c_P$. The shape factor of the crystals and the density of the crystals are given by $k_v$ and $\rho_{\text{cryst}}$. For crystallization phenomena, the growth rate is given by $G$ and the birth and death rates due to agglomeration are given by $B_{\text{Agg}}$ and $D_{\text{Agg}}$. For the mass transferred from the solution to the crystals, the second moment of the distribution $\mu_2$ is necessary. The differential equations for temperature and concentration are solved using an explicit Euler scheme to be consistent with the overall solution method. For the solution of the population balance equation \eqref{eq:PBE} a constant-time step Monte Carlo solution method \cite{vanpeborghgoochMonteCarloSimulation1996} is used. The model should reflect the inherent slug-to-slug variability of the particle size distribution, therefore, no constant $N$ method where the number of particles in the Monte Carlo simulation is kept constant is used. The initialization of the particle size distribution is performed according to the inlet conditions of the slug flow crystallizer. The particles are sampled from a given initial distribution until the mass of the particle size distribution corresponds to the crystal mass at the inlet for the time step:
\begin{align}\label{equation:SlugSize}
    m_{\text{cryst,new}} = w_{\text{cryst}}\dot m_{PM}dt,
\end{align}
where $w_{\text{cryst}}$ corresponds to the crystal mass fraction of the inlet flow. Hence, crystals are sampled until the crystal mass in a slug matches the theoretical crystal mass of a time step, regardless of whether this matches the size of actual slugs. The difference arises because for the duration of a simulated time step $dt$, the number of slugs entering the real slug flow crystallizer does not have to be one. We achieve a connection to the real slug-to-slug variability by reducing the population at the outlet to the number of crystals that would be in a single real slug. 
It should be noted that using this approach, the variability could be calculated for an arbitrary population size or time period.

\begin{figure}
    \centering
    \includegraphics{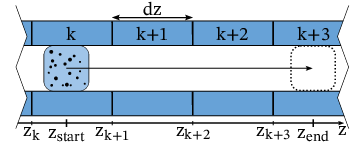}
    \caption{Illustration of a single slug advanced for one time step. Due to the nature of the solution method, it is possible for a slug to pass several finite volumes of the external tempering medium in one time step. This must be taken into account when solving the heat balance.}
    \label{figure:ModelSketchSingleSlug}
\end{figure}

The difference in discretization schemes between the process medium and the tempering medium leads to some difficulties. The heat flow between the process medium and the tempering medium, described in \eqref{eq:TempPM} and part of $s(T)$ in \eqref{eq:PDETemperingMedium} contains the temperature difference $\Delta T=T_{TM}-T_{PM}$. Figure~\ref{figure:ModelSketchSingleSlug} illustrates the difficulties that arise when computing $\Delta T$.
Each outer finite volume corresponds to a single temperature for $T_{TM}$. Since it is possible for a single slug to pass multiple outer finite volumes in one time step, we adapt the computation of the heat flow. After the slug has advanced, the heat flow between the slug and the tempering medium is calculated. For Figure~\ref{figure:ModelSketchSingleSlug} this corresponds to $\Delta T = T_{PM}-T_{TM,k+3}$. The temperature of the slug is updated using this temperature difference. Since it is clear that a pure heat transfer with the $k+3$-rd element and no heat transfer to the $k$-th, $k+1$-st, $k+2$-nd element is not true in reality we compute linearly the contribution to each outer finite volume and use this value for the solution of \eqref{eq:PDETemperingMedium}. Exemplary for Figure~\ref{figure:ModelSketchSingleSlug} this leads to:
\begin{align}
    \dot Q &= U_{PM,TM}A_i(T_{TM,k+3}-T_{PM,i}), \\
    \dot Q_{\text{normalized}} &= \frac{\dot Q}{z_{\text{end}}-z_{\text{start}}}, \\
    \dot Q_k&= \dot Q_{\text{normalized}}(z_{k+1}-z_{\text{start}}), \\
    \dot Q_{k+1}= \dot Q_{k+2}&= \dot Q_{\text{normalized}}dz, \\
    \dot Q_{k+3}&= \dot Q_{\text{normalized}}(z_{\text{start}}-z_{k+3}),
\end{align}
where $\dot Q$ corresponds to the heat flow. Further adaptations and improvements to the linear scheme used to calculate individual contributions are possible.

\begin{figure}
    \centering
    \includegraphics{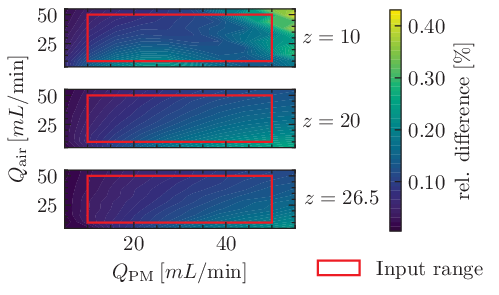}
	\caption{Relative temperature difference at different positions of the slug flow crystallizer during steady-state operation, comparing the presented case (where slugs may not coincide with actual slugs) versus the case using a correlation for the actual slug length from \cite{kufnerModelingContinuousSlug2023}.}
    \label{figure:RelativeErrorNovelModel}
\end{figure}

The novel model employs a computational simplification where a single slug is introduced at each time step with the size of the slug given by:
\begin{align}\label{equation:SlugSize2}
    m_{\text{new}} = dt \, \dot m_{PM},
\end{align}
where $m_{\text{new}}$ is the mass of the slug introduced at the inlet for the respective time step and $\dot m_{PM}$ is the mass flow of the process medium. Although computationally efficient, this approach does not necessarily reflect the actual physical mechanisms governing the formation of slugs in reality. To validate that this modeling decision does not significantly impact accuracy, its effects are investigated through comparative simulations. Since slug size affects heat transfer, and thus supersaturation levels and crystallization phenomena, our simplified approach is compared against a model using physically realistic slug sizes. In reality, the size of the slugs for a given slug flow crystallizer is determined by the flow of the process medium $Q_{\text{PM}}$ and the flow of air $Q_{\text{air}}$. To represent the case where the simulated slugs represent the actual slugs, the slug length correlation from \cite{kufnerModelingContinuousSlug2023} is used to calculate the length of the slugs. For different combinations of process medium and air flow rates, the time step is chosen such that in \eqref{equation:SlugSize2} the slug size coincides with the realistic slug size. Then the simulations are run to steady state. The results obtained for realistic slug sizes are compared with our model using a computationally efficient time step of $dt=5 \, s$ for identical flow conditions. The relative difference between the temperatures at different positions is shown in Figure~\ref{figure:RelativeErrorNovelModel}.
The results show maximum relative temperature differences of less than $0.4 \, \%$ (corresponding to $\sim 1 \, K$) within our operational input space (red box in figure). The overall small error justifies using the proposed computationally efficient approach, where the slug lengths do not necessarily coincide with the lengths of actual slugs. The results show that the simulation time step must be chosen reasonably small to achieve a small error. The investigation of the error is an important part of the model design process, and it can be easily adjusted by accordingly adapting the time step. In any case, a constant time step must be chosen before the simulation. Direct application of the correlation is not possible since in this case the time step would have to be changed during simulation depending on the volume flow which is not possible within a model predictive control scheme.

\begin{figure}
    \centering
    \includegraphics{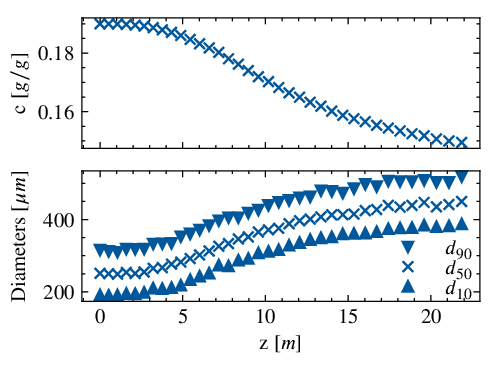}
	\caption{Results of the proposed model for an exemplary simulation. The concentration and some characteristic diameters of the particle size distribution are plotted over the length of the crystallizer at a certain time. The model yields the full distribution at the outlet as well as the temperatures of tempering and process medium.}
    \label{figure:NovelModel}
\end{figure}

The results of an example simulation are shown in Figure~\ref{figure:NovelModel}.
The model was simulated with some arbitrary inputs, and the concentration as well as some characteristic diameters of the particle size distribution are plotted over the length of the crystallizer $z$.
The code for the model as well as the used parameters can be found in our repository\footnote{\nolinkurl{https://github.com/collinj2812/multistage_for_SFC}\label{Github}}.

\section{Surrogate model development}\label{section:SurrogateModels}

The purpose of the model developed in the previous chapter is the usage in MPC. The goal is to control certain process parameters (e.g. characteristic diameters of the particle population) under the presence of uncertainty. Since the first-principle model is very complex and it is not possible to obtain gradient information of the model, it is inherently difficult to optimize. Therefore, we approximate our model to obtain a surrogate model suitable for optimization and MPC. We first gather large data sets from open-loop simulations by exciting the system and subsequently use the data sets to train data-based models. We will focus on data-based models based on neural networks. Data-based models exhibit uncertainty due to the inherent variability of the process, uncertain parameters of the first-principle model, and approximation errors. Since we desire to use the uncertainty within the MPC scheme, we will train models using CQR and BLL neural networks, which give measures of uncertainty of their predictions.
To obtain accurate and efficient models, the models should be as simple as possible and only predict states that are important to the control task. For the slug flow crystallizer we are only interested in the states leaving the system at the outlet. Therefore, we only predict a subset of the states at the outlet which we will call measurements $y$. To overcome observability issues, we use nonlinear autoregressive models with exogenous inputs (NARX) \citep{ljungSystemIdentification2017} for our model predictions. Instead of predicting the states at the next time step given the states and inputs at the current time step, for NARX, the measurements at the next time step are predicted using measurements and inputs at the current time step as well as measurements and inputs at past time steps:
\begin{align}\label{eq:NARX}
    y_{k+1} = f(y_{k},\ldots,y_{k-l},u_{k},\ldots,u_{k-l}),
\end{align}
where $k$ denotes the time step and $u$ represents the inputs of the system. The lag parameter $l$ determines the number of past time steps considered for prediction.

For reasons of notational consistency with literature we call the input of the subsequent data-based models $X$ and outputs $Y$, where, respectively, a single row contains a sample and the columns correspond to the features. Comparison to \eqref{eq:NARX} gives the simple relation for a single sample assuming scalar inputs and measurements:
\begin{align}\label{eq:SingleSample}
    X_i &= (y_{k},\ldots,y_{k-l},u_{k},\ldots,u_{k-l}), \\
    Y_i &= (y_{k+1}).
\end{align}
Given the training data, we train our neural network model $\text{NN}$ to predict:
\begin{align}\label{eq:Y=f(X)}
    Y=\text{NN}(X).
\end{align}
As the baseline data-based model we train a standard feedforward neural network which consists of multiple layers containing each a linear and a nonlinear transformation. The computation for the $i$-th hidden layer is given by:
\begin{align}\label{eq:Activation}
    a_{i+1}=h(w_{i}^Ta_i),
\end{align}
where $a$ is the activation. The activation for the first hidden layer consists of the inputs. The activation function is given by $h$ and can be freely chosen, although, in regression, usually a linear activation function is chosen for the last layer. The trainable weights of the neural network are given by $w$. To train the network, a loss function is used to measure the quality of the predictions. For the standard feedforward neural network, we use the mean squared error (MSE). The optimization problem to train the neural network using the MSE is given as:
\begin{subequations}\label{eq:OptProblemNeuralNetwork}
\begin{alignat}{2}
    \min_{w} \ \frac{1}{N}\sum^N_i\left|Y_i-\hat Y_i \right|_2^2,
\end{alignat}
\end{subequations}
where $\hat Y_i$ represents the prediction for sample $i$. For subsequent models, specialized loss functions will be necessary.
 
\subsection{Conformalized quantile regression}

As the first method to obtain a data-based model that can quantify the uncertainty of its predictions, we use conformalized quantile regression \citep{romanoConformalizedQuantileRegression2019}. Here, as before, a neural network is trained to predict the next measurement. In addition, two neural networks are trained to predict quantiles for the next measurement. Then, the quantile models are conformalized. This means that the prediction of the quantile models is corrected by a fixed value to adhere to the predetermined quantile for a predetermined probability using previously unseen calibration data. The loss function differs from before since the objective has changed. To obtain a quality of fit for the quantiles, the pinball loss function is used \citep{romanoConformalizedQuantileRegression2019}. The loss function for a single sample is given by:
\begin{align}\label{eq:PinballLoss}
    L_\alpha(Y_i,\hat{Y}_i):= \begin{cases} 
                                \frac{\alpha}{2}(Y_i-\hat{Y}_i) & \text{if } Y_i-\hat{Y}_i > 0,\\
                                (1-\frac{\alpha}{2})(\hat{Y}_i-Y_i), & \text{otherwise}
\end{cases}
\end{align}
where $Y_i$ is the true value from the training data and $\hat{Y}_i$ is the predicted value. Using this loss function will lead to a model that predicts the $\alpha/2$-th quantile. The parameter $\alpha$ is called the miscoverage level. For the prediction of the mean, that is, the $50$-th quantile, \eqref{eq:PinballLoss} corresponds to the mean absolute error (MAE).

Subsequently, the quantile models are conformalized. Each model predicts all samples from the unseen calibration data set, and a conformity score $E_i$ is calculated:
\begin{align}\label{eq:ConformityScore}
    E_i = \text{max}\left\{\text{NN}_{\text{lo}}(X_i)-Y_i,Y_i-\text{NN}_{\text{up}}(X_i)\right\},
\end{align}
where $\text{NN}_{\text{up}}$ and $\text{NN}_{\text{lo}}$ correspond to the upper and lower quantile models. Finally, the $(1-\alpha)$-th empirical quantile of the conformity scores is determined which will serve as the fixed offset $Q_{1-\alpha}$ added to the upper quantile prediction and subtracted from the lower prediction. The final prediction interval for a sample $X_i$ is given by:
\begin{align}\label{eq:IntervalPredictionCQR}
    \left[\hat{Y}_{i,\text{lo}},\hat{Y}_{i,\text{up}}\right] = \left[\text{NN}_{\text{lo}}(X_i)-Q_{1-\alpha}, \, \text{NN}_{\text{up}}(X_i)+Q_{1-\alpha}\right].
\end{align}
The conformalization step is necessary to ensure the coverage level of $\alpha$ on unseen test data. The prediction interval is adjusted using a constant offset, such that a coverage of $\alpha$ is achieved on an independent data set that has not been used before for training or validation (calibration data set). Consequently, this leads to statistical guarantees for coverage on exchangeable test data (joint probability distribution is invariant under permutations) \cite{romanoConformalizedQuantileRegression2019}.

\subsection{Bayesian last layer neural networks}

Bayesian last layer neural networks \citep{fiedlerImprovedUncertaintyQuantification2023a, lazaro-gredillaMarginalizedNeuralNetwork2010, watsonLatentDerivativeBayesian2021} represent a fundamentally different way of quantifying prediction uncertainty compared to CQR. The neural network returns Gaussian distributed predictions where the variance of the predictions can be interpreted as a measure of uncertainty, with high variance corresponding to high uncertainty and small variance corresponding to low uncertainty.

The setting for BLLs is to find the function $f$ that generated a data set $(X,Y)$ given some additive white Gaussian noise:
\begin{align}\label{eq:RegressionSetting}
    X_i&=f(Y_i)+\epsilon_i, \\
    \epsilon&\sim \mathcal{N}(0,\beta_\epsilon),
\end{align}
where $X_i$ and $Y_i$ again represent single samples of the training data set. In contrast to full Bayesian neural networks, we assume only the weights of the last layer to be Gaussian distributed. We choose a linear activation function for the last layer which leads to the following relation for the prediction of the BLL neural network:
\begin{align}\label{eq:WeightsLastLayer}
    \hat{Y_i}=w_{n+1}^T\Phi(X_i),
\end{align}
where $w_{n+1}$ corresponds to the weights of the last layer which are Gaussian distributed. The activation $a_i=\Phi(X_i)$ is the output of the $n$-th, and therefore the last hidden layer. The weights of the last layer $w_{n+1}$ are determined by computing their posterior distribution given the data and a prior noise covariance $\beta_w$ for the weights using Bayes rule. The free parameters are then the weights of the hidden layers $w_0,\ldots,w_n$, as well as the noise covariances $\beta_\epsilon$ and $\beta_w$. The parameters are determined by maximizing the log marginal likelihood function which corresponds to the denominator in Bayes rule. Maximizing the log marginal likelihood function is common practice in a Bayesian setting and leads to an approximation of a full Bayesian neural network \citep{fiedlerImprovedUncertaintyQuantification2023a}. Since the weights of the last layer are already given by \eqref{eq:WeightsLastLayer}, computationally demanding equality constraints are necessary for model training. Fortunately, it has recently been shown that when using the log marginal likelihood as a loss function for model training, the equality constraint for the weights of the last layer can be neglected \citep{fiedlerImprovedUncertaintyQuantification2023a}, leading to computationally efficient training of BLL neural networks.

BLL models provide Gaussian distributions as prediction. By adding and subtracting multiples of the standard distribution we can generate a prediction interval similar to the interval for CQR. We can adjust the width of the interval by changing the multiple $m$:
\begin{align}\label{eq:IntervalPredictionBLL}
    \left[\hat{Y}_{i,\text{lo}},\hat{Y}_{i,\text{up}}\right] = \left[\mu_i-m\sigma_i, \, \mu_i+m\sigma_i\right],
\end{align}
where $\mu_i$ represents the mean of the prediction $f_{\text{BLL}}(X_i)$ and $\sigma_i$ represents the standard deviation.

\subsection{Comparison of surrogate models}

We investigate the differences between the different modeling approaches by comparing the prediction on unseen test data. The models were trained on a training data set containing $50 \, 000$ training samples. The model architectures were chosen equal. For each model, one hidden layer was used with $30$ neurons. For the prediction of the quantile models $10$ neurons were used. As activation function the GELU function \citep{hendrycksGaussianErrorLinear2023} was used. For all models, the lag parameter for the NARX structure was chosen to be $l=4$, with a time step of $50$ seconds. The CQR models were obtained using $\alpha=0.05$. For the BLL model, we compute the uncertainty by adding $\pm 2\sigma$ to the mean prediction. The tool do-mpc \citep{fiedlerDompcFAIRNonlinear2023} with CASADi \citep{anderssonCasADiSoftwareFramework2019} was used to implement the models. For the training of the standard neural network and the CQR model PyTorch was used \citep{Ansel_PyTorch_2_Faster_2024}. For the BLL model the implementation from \cite{fiedlerImprovedUncertaintyQuantification2023a} with Keras \cite{chollet2015keras} and Tensorflow \citep{tensorflow2015-whitepaper} was used. The code for the results can be found in our repository\footref{Github}.

In the investigated case the uncertainty in the process, i.e. the process variability, stems from the sampling of the initial distribution. The proposed data-based models can quantify the process variability, as well as the uncertainty due to lack of data. Generally, it is possible to also consider further uncertainties, such as parametric uncertainties in the model or additive noise.

\begin{figure*}
    \centering
    \includegraphics{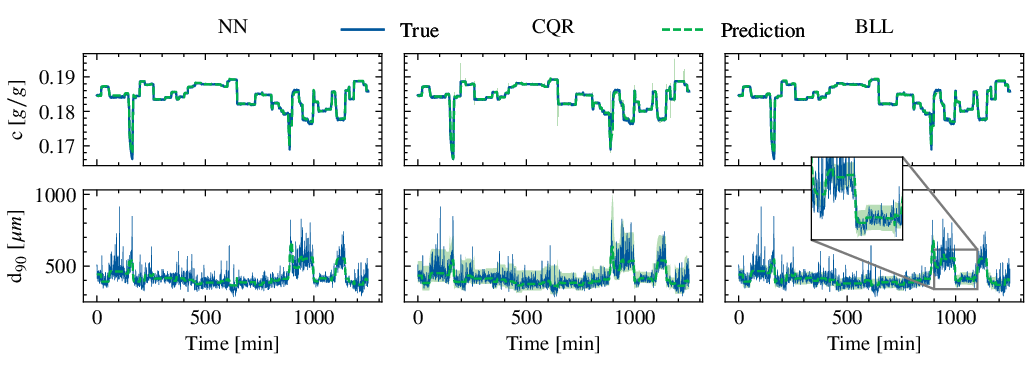}
	\caption{Prediction of the different models on the same sequence of unseen test data. The results are obtained by evaluating the models as simulation models as shown in \cite{schoukensNonlinearSystemIdentification2019}. For a given input sequence and an initial state the model is evaluated by recursively using the output of the model as the input for the next time step.}
    \label{figure:SurrogateModels}
\end{figure*}

The results for the trained models are summarized in Table~\ref{tab:ResultsSurrogateModels}. The inputs to generate the test data were chosen in the same manner as for the training data to randomly excite the system and explore the state space.
\begin{table}[h]
    \centering
    \begin{tabular}{cccc}
        \toprule
         & NN & CQR & BLL \\
        \midrule
        MSE [-] & $1.69 \cdot 10^{-3}$ & $7.40 \cdot 10^{-4}$ & $9.10 \cdot 10^{-4}$ \\
        Coverage [\%] & $-$ & $95.08$ & $96.01$ \\
        \bottomrule
    \end{tabular}
    \caption{Results for the different surrogate models on unseen test data. The models are evaluated as prediction models as shown in \cite{schoukensNonlinearSystemIdentification2019}. The models are tested directly on the test data and compared to the respective label.}
    \label{tab:ResultsSurrogateModels}
\end{table}
All models can provide very accurate results. The CQR and BLL models yield slightly more accurate results than the standard neural network. The target coverage  for CQR ($95\%$ because $\alpha=0.05$) is achieved very accurately. The BLL model assumes Gaussian distributed states, where the chosen interval of $\pm 2\sigma$ would lead to a coverage of $95.45\%$, which also fits very well to the coverage achieved for the test data. Figure~\ref{figure:SurrogateModels} shows the results for unseen test data for concentration of the liquid phase and characteristic diameter $d_{90}$ of the particle size distribution at the crystallizer outlet over time. The blue curves represent the results of the first-principle model as true values. The results for the different surrogate models (NN, CQR and BLL) are given as green dashed lines.
The standard neural network (left) can accurately predict the mean of the states. The model does not provide uncertainty quantification of the predictions. The CQR model (middle) can also accurately predict the mean of the test data, and the model also provides information on the uncertainty of its prediction (green shaded area). Especially for the lower plot of $d_{90}$, it can be seen that the uncertainty quantification of the prediction fits the process variability very well. The predictions are adaptive, indicating high certainty for regions with small process variability and high uncertainty for regions with larger process variability. Also, the prediction interval does not need to be symmetric around the mean prediction, which fits the variability here especially well. The BLL model (right) can also accurately predict the mean of the test data. The predicted uncertainty of the model on the other side is not adaptive. The reasoning behind this is based on the way in which the BLL model obtains the uncertainty quantification of its predictions. The uncertainty quantified by the BLL models is split into two parts. The uncertainty in the weights leads to the indicated prediction uncertainty being adaptive to the degree of extrapolation. In addition, the uncertainty which is due to the additive uncertainty $\epsilon$ of the data as in \eqref{eq:RegressionSetting} is added to the predictions. By choosing a specific value to obtain a prediction interval (in this case $\pm 2\sigma$), we only add and subtract a constant value to our predictions if the uncertainty in the weights is very small. This is the case for the given investigation. Since a large data set was used for model training, the BLL method could very accurately determine the underlying model of the data $f$ from \eqref{eq:RegressionSetting}. Accordingly, the uncertainty indicated due to approximation errors is nearly zero. This leads to an uncertainty description similar to that of purely conformalizing the mean as in CQR. Therefore, the predicted uncertainty is not adaptive to the inherent variability of the process.

\section{Model predictive control}\label{section:MPC}

In model predictive control a model is used to compute optimal inputs for a given prediction horizon $N_{\text{pred}}$. The control goals are formulated in an objective function. The inputs of the system are subsequently determined such that the objective function is minimized for the length of the prediction horizon. 
The MPC problem is repeatedly solved at each time step, where only the first input of the sequence is applied to the system. The MPC optimization problem is given by:
\begin{subequations}\label{eq:MPC}
\begin{alignat}{2}
    &\min_{u_k} &\quad &\sum_{k=0}^{N_{\text{pred}}-1} l(s_k, u_k) + V_f(s_{N_{\text{pred}}}) \\
    &\, \, \, \text{s.t.} & & s_{k+1} = f(s_k, u_k) \\
    &   & &g(s_k,u_k) \leq 0 \\
    &   & &s_0 = s_{\text{initial}},
\end{alignat}
\end{subequations}
where $s_k$ and $u_k$ represent the states and inputs at time step $k$. The objective function consists of the stage cost $l(s_k, u_k)$, which can be a function of the states and inputs, and the terminal cost $V_f(s_{N_{\text{pred}}})$, which is a function of the states. Furthermore, we can enforce state constraints and input constraints by $g(s_k,u_k)$. As a final constraint for the optimization problem, the state trajectory must start at $s_{\text{initial}}$ which is the state of the system at the respective time step. We assume that the states used for the NARX model are measured directly. The measured states are the temperatures $T_{\text{PM}}$ and $T_{\text{TM}}$ at the outlet, the concentration $c_{\text{PM}}$ at the outlet, as well as the three characteristic diameters of the particle size distribution $d_{10}$, $d_{50}$, and $d_{90}$ at the outlet. For a real implementation, an observer could be designed. To focus on the analysis of the MPC performance, we assume direct measurement.

\begin{figure}
    \centering
    \includegraphics{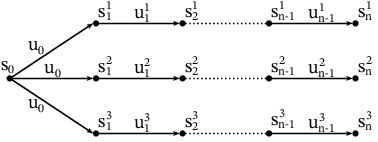}
    \caption{Sketch of the branching performed in multi-stage MPC \citep{luciaMultistageNonlinearModel2013}. Here, the branching is performed once, followed by regular MPC for each branch.}
    \label{figure:MultiStageSketch}
\end{figure}

As internal model for the MPC problem in \eqref{eq:MPC} we use the derived surrogate models from Section~\ref{section:SurrogateModels}. To utilize the uncertainty information of the CQR and BLL models, we propose to use a multi-stage MPC scheme as proposed in \cite{luciaMultistageNonlinearModel2013}. In its original form, multi-stage MPC formulates a scenario tree where each branch of the tree represents a possible value of the uncertainty. This typically represents different possible values of uncertain parameters.
A sketch can be seen in Figure~\ref{figure:MultiStageSketch}. An uncertain parameter is identified where bounds of the parameter are known. 
A weighted sum of all realizations can be used as the objective function of the optimization problem. Constraint satisfaction is enforced for all scenarios.

To integrate the uncertainty quantification of CQR and BLL, we adapt the original multi-stage MPC formulation. Instead of single uncertain parameters in the first-principle models, the predictions of the model are uncertain. We propose to employ the uncertainty in the multi-stage scheme by explicitly computing uncertain scenarios based on the predicted uncertainty of the model, as previously shown in our previous work \cite{johnsonRobustNonlinearModel2024}. To avoid considering uncertainty in each of the states of the system model, we identify important states that are subject to constraints. We consider the uncertainty quantification of these important states to define the branches of the scenario tree of the proposed multi-stage MPC scheme.
By enforcing constraint satisfaction for the uncertain scenarios, the algorithm chooses the back-off from the constraint adaptively based on the uncertainty of the predictions.

For the CQR models the branching is performed using the models predicting the quantiles. We define the NARX state at time step $k$ as $s_{\text{NARX},k}=(s_{k},\ldots,s_{k-l},u_{k},\ldots,u_{k-l})$. The branching is performed as follows:
\begin{align}\label{eq:CQRInMultiStage}
     \left(\begin{array}{c}
         s_{k+1}^1 \\
         s_{k+1}^2 \\
         s_{k+1}^3
    \end{array} \right) =
    \left(\begin{array}{c}
         \text{NN}_\text{up}(s_{\text{NARX},k})+Q_{1-\alpha} \\
         \text{NN}_\text{MAE}(s_{\text{NARX},k}) \\
         \text{NN}_\text{lo}(s_{\text{NARX},k})-Q_{1-\alpha}
    \end{array} \right),
\end{align}
where $s_{k+1}^1, s_{k+1}^2, s_{k+1}^3$ denote the states leading to a different branch of the tree, as depicted in Figure~\ref{figure:MultiStageSketch}. 
It is important to note that the NARX state changes for the prediction horizon given the respective branch. For the upper branch for example at time step $k+2$, we compute the NARX state as: 
\begin{align}
    s_{\text{NARX},k+2}^1&=(s_{k+2}^1,s_{k+1}^1,s_{k},\ldots,s_{k-l+2}, \\
    & \quad \quad \quad \quad  u_{k+2}^1,u_{k+1}^1,u_{k},\ldots,u_{k-l+2}). \nonumber
\end{align}
We construct the multi-stage scheme for the BLL model as described in our previous work \cite{johnsonRobustNonlinearModel2024}. The uncertain branches are computed by adding or subtracting the standard deviation to the mean of the prediction of the BLL neural network. For a prediction of our BLL neural network $\mathcal N(\mu_k, \sigma^2_k)=f_{\text{BLL}}(s_{\text{NARX},k})$, we compute:
\begin{align}\label{eq:BLLInMultiStage}
     \left(\begin{array}{c}
         s_{k+1}^1 \\
         s_{k+1}^2 \\
         s_{k+1}^3
    \end{array} \right) =
    \left(\begin{array}{c}
         \mu_k+m\sigma_k \\
         \mu_k \\
         \mu_k-m\sigma_k
    \end{array} \right),
\end{align}
where $m$ is a tuning parameter which can be freely chosen. By choosing $m$, we can adjust the level of conservatism of the controller. A large value for $m$ will lead to a conservative performance with a larger back-off from a constraint. Smaller values will lead to the controller being less conservative, going closer to the constraint with a higher risk of closed-loop constraint violations. For the CQR model, this tradeoff can be influenced by changing $\alpha$.

\section{Simulation results}\label{section:Results}

To show the differences between the different models within an MPC algorithm, we choose different scenarios that illustrate the strengths and weaknesses of the models. For the simulator, acting as the real system, we choose our detailed but not optimizable model from Section~\ref{section:Model}. As internal model, we choose the data-based models shown in Section~\ref{section:SurrogateModels}. For the data-based models we use our multi-stage MPC scheme from \ref{section:MPC} for the CQR as well as for the BLL model. The standard feedforward neural network model is used directly in the optimization problem \eqref{eq:MPC}. Code for all results is openly available\footref{Github}.

The chemical system used for the investigation is L-alanine/water as presented in \cite{termuhlenContinuousSlugFlow2021}. The used parameters and correlations for the simulation studies are shown in Appendix~\ref{appendixsection}.

\subsection{Control goals}\label{subsection:ControlGoals}

The objective of the controller is to maximize the amount of produced crystals as well as the size of the crystals, which is a common objective in crystallization. Relating our goals to our system models, we want to maximize the flow rate of the process medium $Q_{PM}$ as well as our median particle diameter $d_{50}$. In addition, the flow rate of the tempering medium, i.e. the cooling liquid is minimized. For subsequent downstream processes, very large crystals can be problematic. Therefore, we enforce a constraint as an upper bound on the characteristic diameter $d_{90}$. Our cost function for the MPC problem \eqref{eq:MPC} is, therefore, given as:

\begin{align}\label{eq:CostFunction}
    &l(s_k,u_k)= - \gamma_1 d_{50,k} -\gamma_2Q_{PM,k} +  \gamma_3Q_{TM,k} + \gamma_4\Delta u  , \\
    &V_f(s_{N_{\text{pred}}})=-\gamma_1d_{50,{N_{\text{pred}}}},
\end{align}
where the parameters $\gamma_i$ represent weighting factors of the cost function. The manipulated variables are the process medium flow rate $Q_{PM}$, the gas flow rate $Q_{air}$, as well as the flow of the tempering medium $Q_{TM}$. The state constraints consist only of an upper bound for $d_{90}$ which are implemented as a soft constraint. The input constraints are chosen as box constraints coinciding with the input ranges used for data generation in Section \ref{section:SurrogateModels}.

\subsection{Case studies}\label{subsection:CaseStudies}

We consider two different case studies to investigate the two different aspects of the proposed algorithm  when controlling the continuous slug flow crystallizer model developed in Section~\ref{section:Model}. For the first case study, we use a large training data set to keep approximation errors to a minimum. For the second case study, we investigate how our proposed algorithm performs for varying approximation errors. Hence, we use smaller data sets of varying size, leading to solutions also based on extrapolated predictions. The developed data-based models exhibit uncertainty due to inherent process variability, uncertain parameters in the first-principle models, and also approximation errors. In the scope of this work, we focus on the investigation of uncertainty due to process variability and approximation errors. Process variability is present in our model especially in the particle size distribution. Using a Monte Carlo method for the solution of the population balance equation leads to a high variability that is close to reality. Consequently, the process variability will always be present in our solutions. 

\begin{figure*}
    \centering
    \includegraphics{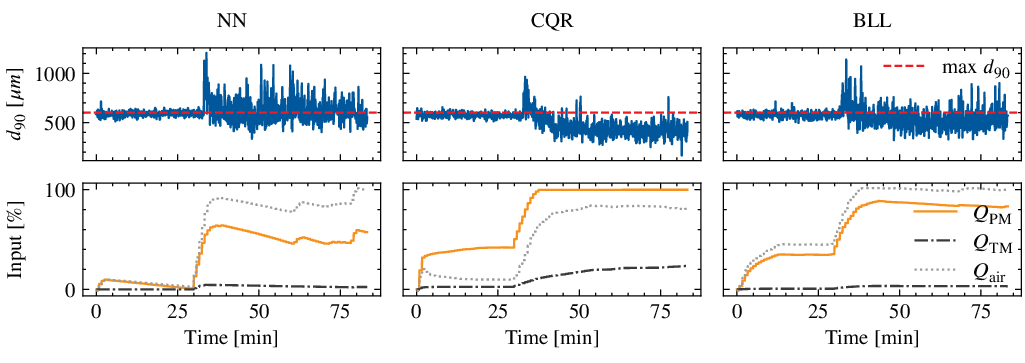}
	\caption{Comparison of different models used in MPC with the cost function from \eqref{eq:CostFunction}. For the inputs, the orange full line represents the volume flow of the process medium, the dotted gray line represents the volume flow of air, and the dashdotted black line represents the volume flow of the tempering medium.}
    \label{figure:MPCdifferentModelsHeteroscedasticityAnalysis}
\end{figure*}

For the first case study, we generate a training data set with $50 \, 000$ data points and train a standard feedforward neural network, a CQR model, and a BLL model. All CQR models were obtained using $\alpha=0.05$. For the BLL models $m=2$ was chosen. The cost function from \eqref{eq:CostFunction} is used for the MPC problem. The scenario for all three models is the same. The first-principle model acting as the simulator is simulated into a steady-state. Subsequently, the controller is turned on. After $35$ time steps ($\approx 29.17$ min) the weight fraction of the crystals at the inlet of the crystallizer is changed from $w_{\text{cryst}}=0.01$ to $w_{\text{cryst}}=0.001$, leading to a significantly higher variability of the constrained state $d_{90}$. The results using these models are shown in Figure~\ref{figure:MPCdifferentModelsHeteroscedasticityAnalysis}. Key values of the case study are shown in Table~\ref{tab:ResultsFirstCaseStudy}.
The standard neural network does not offer uncertainty quantification of its predictions. Consequently, the constraint on $d_{90}$ acts only on the mean of the state. The variation in the process leads to significant violations of the constraint. For the simulation performed, the model predictive controller using the standard neural network violated the constraints in more than $30 \, \%$ of the time steps. 

The controller using the CQR model with the proposed scheme of ~\eqref{eq:CQRInMultiStage} adapts to the sudden change in the variability of the process. After violating the constraints  with the sudden change in $w_{\text{crystal}}$, the method adapts to the larger variability of the process and adequately increases the back-off from the constraint. The method violates the constraint only in $8 \, \%$ of the time steps, mainly at the change in $w_{\text{crystal}}$. As seen in Figure \ref{figure:SurrogateModels}, the method using the BLL model cannot adapt to the change in process variability. Using the BLL model as proposed in \eqref{eq:BLLInMultiStage} illustrates the lack of adaptability. By acting like a conformalization step, the controller keeps a constant back-off from the constraint, which is not adapted dynamically. The average constraint violation, which is computed over all time steps of the respective simulation, is also the lowest for the simulation using the CQR model. The average constraint violation for the simulation using BLL is higher, but compared to the standard neural network, the usage of the BLL models uncertainty leads to a better performance. For both CQR and BLL, the number of constraint violations and the magnitude of violations can be reduced by changing $\alpha$ or $m$. The cost per time step achieved is best for the standard neural network at the price of increased constraint violations.
\begin{table}[h]
    \centering
    \begin{tabular}{ccccc}
        \toprule
         & \multicolumn{2}{c}{Constraint violation} & Avg cost per & Avg CPU \\
         & \% of steps & avg rel & time step & time \\
        \midrule
        NN & $31$ & $3.7 \, \%$ & $-5.28$ & $2.1 \, s$ \\
        CQR & $8$ & $1.0 \, \%$ & $-4.25$ & $3.8 \, s$ \\
        BLL & $32$ & $2.4 \, \%$ & $-4.88$ & $3.8 \, s$ \\
        \bottomrule
    \end{tabular}
    \caption{Results of the first case study using a large training data set. Constraint violations and cost are calculated at the time steps of the controller. The average relative constraint violations are computed over all time steps of the simulation. The cost average is the scaled cost not considering the cost of the soft constraint and penalty terms on the change of the inputs.}
    \label{tab:ResultsFirstCaseStudy}
\end{table}
The computation time for all algorithms was found to be capable of a real-time application with the maximum time to solve the MPC problem for all methods being less than $6$ seconds. The average time to solve the MPC problem was fastest for the neural network where no branching was performed. The CPU times for CQR and BLL were equally fast.

\begin{figure*}
    \centering
    \includegraphics{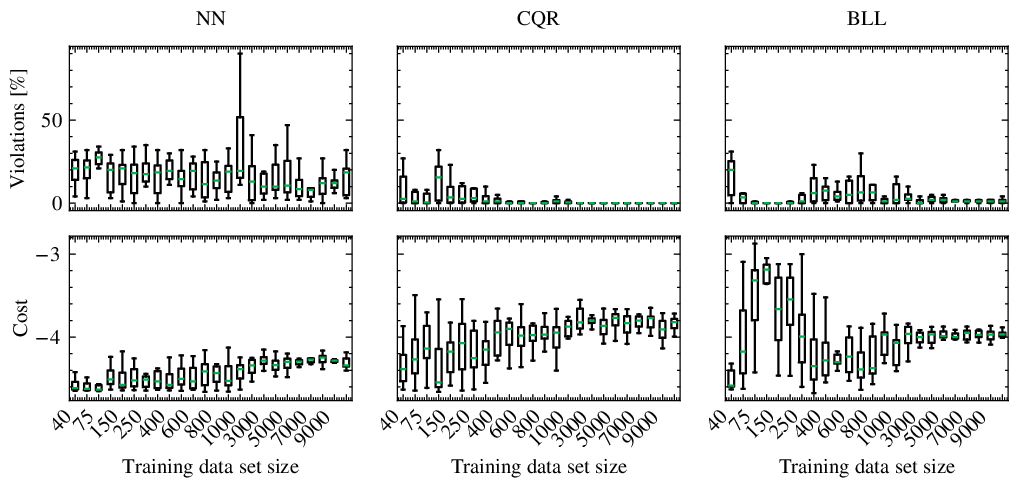}
	\caption{Analysis of the MPC performance using data sets of different sizes for the different surrogate models. For each data set size $10$ different data sets as well as $10$ different models where trained and compared. The results for each size are presented as boxplots. The green line within the box represents the median of the respective simulation. The box itself represents the first and third quartiles. The box plot is extended to the horizontal line on each side adding $1.5$ times the inter-quartile range to the box. The top plots show the percentage of time steps where constraint violations occurred. The bottom plot shows the attained closed-loop cost for the MPC runs. The cost is computed without taking the penalization of the change in inputs and the soft constraint into consideration.}
    \label{figure:MPCdifferentModelsDatsetAnalysis}
\end{figure*}

For the investigation of the second case study, we aim to analyze the performance of the different models in the MPC scheme when the approximation error of the models is not negligible. The approximation error increases when less data is available for model training. We compare the performance using data sets of different sizes for model training. To reduce random effects in small data sets, we generate $10$ data sets for each investigated size. Then, we train a standard neural network, a CQR model, and a BLL model on each data set. In total, we investigated data sets of $24$ different sizes, leading to $240$ data sets and $720$ different models. Figure~\ref{figure:MPCdifferentModelsDatsetAnalysis} shows the results for the different models.
For the standard neural network, the model performance becomes more consistent for larger data sets. The cost, which is desired to be minimized, of the different MPC runs converges and becomes more consistent. However, the increasing amount of data cannot lead to less constraint violations in the case of the neural network because of the lack of uncertainty quantification.

The performance of the CQR models improves significantly for larger data sets. The models can consistently satisfy the constraints. The BLL model also has better performance for larger data sets and leads to clearly better results than using the standard neural network directly in the MPC algorithm. In comparison to the CQR model, the performance of the BLL model is slightly worse in terms of constraint violations because of the lack of adaptation in the uncertainty quantification.
We believe that the uncertainty quantification capabilities of CQR and BLL can lead to satisfactory performance of MPC controllers even when it is not possible to gather large amounts of data for the surrogate models. We expect this to be especially relevant for complex large-scale systems.

\subsection{Discussion on the operational advantages of the proposed approach}

The slug flow crystallizer under investigation presents unique challenges that render traditional control strategies inadequate. PID controllers, for example, are inherently not suited for highly nonlinear multiple-input multiple-output systems exhibiting large time delays as in the present case. In addition, we aim to use economic cost functions, i.e. we maximize crystal diameters and product streams and minimize cooling flows, while enforcing constraint satisfaction. Traditional control strategies struggle with the given complexity in the presented case. Additionally, the discrete nature of the presented model (discrete slugs and Monte Carlo population) renders the model to be inherently non-optimizable. It is therefore necessary to resort to approximation techniques, e.g. surrogate modeling, to use the model in optimization-based control. For the case studies shown in Section \ref{subsection:CaseStudies}, a comparison to traditional control strategies is therefore not possible and the simplest baseline control strategy is an MPC controller that uses a standard feedforward neural network as internal model as shown on the left in Figure~\ref{figure:MPCdifferentModelsHeteroscedasticityAnalysis}.

The proposed approach is especially interesting for the practical operation of the slug flow crystallizer. Until now, operation of the slug flow crystallizer relies heavily on expert knowledge. This approach has two main drawbacks. First, it is difficult to transfer expert knowledge to new chemical systems or slug flow crystallizer geometries. Second, the slug flow crystallizer is prone to fouling. Therefore, cleaning cycles are part of normal operation. Frequent shut-down and start-up of the process leads to dynamic operation being important to the process.

\begin{figure}
    \centering
    \includegraphics{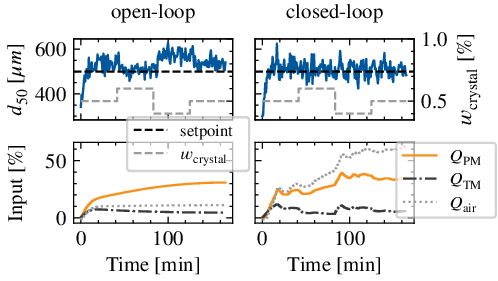}
	\caption{Comparison between open-loop (left) and closed-loop (right) control of the slug flow crystallizer. The crystal mass fraction which acts as a disturbance is plotted in the top plots by the gray dashed line. The blue line in the top plots represents the characteristic particle size diameter $d_{50}$.}
    \label{figure:ComparisonOpenLoopClosedLoop}
\end{figure}

To show the advantages of closed-loop control for the slug flow crystallizer, we present the comparison to open-loop control. The crystal mass fraction at the inlet of the slug flow crystallizer is assumed to be uncertain but measurable and varies along the simulation. The objective is to track a median particle size diameter ($d_{50}=500 \, \mu m$) while also maximizing feed and minimizing cooling flow. For the open-loop controller, we solve the MPC optimization problem for the full simulation once and apply the full computed input trajectory. The closed-loop comparison will be the MPC controller as described in Section \ref{section:MPC} using the standard feedforward neural network as internal model. The same stage cost and terminal cost in the cost function as well as constraints are used for both optimization problems. Figure~\ref{figure:ComparisonOpenLoopClosedLoop} shows the comparison of the trajectories.
Closed-loop control leads to significantly better performance tracking the median diameter. The desired median particle size is consistently achieved by incorporating the measured disturbance of $w_{\text{crystal}}$. The open-loop controller where the input trajectory is computed once at the beginning of the simulation cannot compensate for the disturbance. Note that the particle populations are sampled randomly at each time step. Therefore, the initial population may differ between the open-loop and closed-loop simulation.

The proposed approach will be demonstrated on the slug flow crystallizer in future experimental studies. Fouling detection and automated cleaning will be part of the optimization problem, showcasing the practical advantages of the proposed algorithm controlling the particle size distribution during predominantly dynamic operation.

\section{Conclusion}\label{section:Conclusion}

The development of continuous processes often leads to distributed systems. The slug flow crystallizer is a system that is distributed in spatial direction and particle size distribution. Model-based control using MPC is not directly possible due to the resulting complexity of the models. In this work, we first developed a new dynamic model for the slug flow crystallizer. The presented model addresses the main challenges that make modeling of the slug flow crystallizer difficult. The complete absence of backmixing and a change in velocity along the crystallizer are captured by using an adaptation of the sequencing method. The high and varying process variability is captured by using Monte Carlo simulations to solve the population balance equation, giving a measure of slug-to-slug variability. Data-based models trained with data generated by open-loop evaluations of the first-principle model enable use in the model-based controller. While standard control of the slug flow crystallizer is usually performed using expert knowledge in an open-loop fashion the proposed approach enables optimization-based control.

To account for both approximation errors and process variability in the controller despite the use of a surrogate model, we use conformalized quantile regression (CQR) and Bayesian last layer (BLL) neural network models. We illustrate the advantages and disadvantages of the models in two different case studies. The method using the BLL model is not able to adaptively quantify and account for process variability, but the approximation error due to extrapolation is quantified and can be accounted for in the controller. The controller with the CQR model, on the other hand, cannot adaptively quantify the approximation error according to the degree of extrapolation, but can dynamically quantify and take into account the process variability. The process variability is particularly important for the slug flow crystallizer. Since the variability also varies depending on process parameters, the presented method using CQR models in MPC and directly considering the changing process variability in the controller leads to efficient control of the slug flow crystallizer. The controller can act on changes in process variability and therefore drive the process appropriately close to constraints. The decision as to which model is more suitable is therefore a question of the size of the training data set and the extent of process variability. If the inherent process variability is large and a lot of data is available, CQR offers advantages. If quantification of the approximation error is more important due to insufficient data, an algorithm with BLL will deliver better results.

Future work will study the experimental validation of the proposed approach on the real slug flow crystallizer, which is already in operation \citep{termuhlenContinuousSlugFlow2021}. The proposed model as well as the proposed control techniques will be used to operate the slug flow crystallizer optimally, controlling the uncertain particle size distribution while enforcing constraints.

\appendix\label{appendix}

\section{Supplementary model information}\label{appendixsection}

\begin{table}[pos=h]
\centering
\begin{tabular}{@{}ll@{}}
\toprule
Parameter & Value \\
\midrule
Agg. param$^1$ $\beta_0$ & $2 \times 10^{4}$ \\
Mean init. distribution $\mu_{\text{init}}$ & $2.5 \times 10^{-4}$ m \\
Std  init. distribution $\sigma_{\text{init}}$ & $1.0 \times 10^{-4}$ m \\
Shape factor $k_v$ & $\pi/6$ \\
Crystal density $\rho_{cryst}$ \cite{GESTISSubstanceDatabase} & $1432$ kg/m³ \\
PM density $\rho_{PM}$ & $1000$ kg/m³ \\
PM specific heat $c_{p,PM}$ & $4186$ J/(kg·K) \\
PM-TM heat transfer$^1$ $U_{PM,TM}$ & $9.25 \times 10^{2}$ W/(m²·K) \\
TM density $\rho_{TM}$ & $1000$ kg/m³ \\
TM specific heat $c_{p,TM}$ & $4186$ J/(kg·K) \\
TM-envir. heat transfer$^1$ $U_{TM,env}$ & $8.27$ W/(m²·K) \\
SFC length $L$ & $24$ m \\
PM inner diameter $d_{i,PM}$ \cite{termuhlenQuantificationEvaluationOperating2021} & $3.18 \times 10^{-3}$ m \\
PM outer diameter $d_{a,PM}$ & $4.76 \times 10^{-3}$ m \\
TM inner diameter $D_{i,TM}$ & $1.5 \times 10^{-2}$ m \\
TM outer diameter $D_{a,TM}$ & $1.9 \times 10^{-2}$ m \\
Outlet pressure $p_{out}$ & $1.01 \times 10^{5}$ Pa \\
\bottomrule
\multicolumn{2}{@{}l@{}}{$^1$ fitted to experiments from \cite{termuhlenCharacterizationSlugFormation2019, termuhlenDesignOperationContinuous2022} and \cite{kufnerPredictionParticleSuspension2023}.}
\end{tabular}
\caption{Model parameters for the system L-alanine and water.}
\label{tab:parameters}
\end{table}

\begin{table}[pos=h]
\centering
\begin{tabular}{@{}ll@{}}
\toprule
Correlation & Expression \\
\midrule
Growth rate \cite{hohmannAnalysisCrystalSize2018} & $G=5.857 \times 10^{-5} \Delta S^2 \text{tanh} \left(\frac{0.913}{\Delta S} \right) $ \\
Solubility \cite{wohlgemuthInducedNucleationProcesses2012} & $c^*=0.11238 e^{9.0849 \times 10^{-3} T}$ \\
Supersaturation & $\Delta S=\frac{c-c^*}{c^*}$ \\
Agg. kernel$^1$ & $\beta = \beta_0 G^{\beta_1} v^{\beta_2} \quad (\beta \neq f(L))$ \\
 & $\text{with} \, \beta_0=2 \times 10^4, \beta_1=1, \beta_2 =1$ \\
Init. distribution & $\mathcal{N}(250,100)$ \\
\bottomrule
\multicolumn{2}{@{}l@{}}{$^1$ parameters $\beta_0$,$\beta_1$,$\beta_2$ must be fitted to experiments.}
\end{tabular}
\caption{Model correlations for the system L-alanine and water.}
\label{tab:correlations}
\end{table}

\section*{Declaration of Generative AI and AI-assisted technologies in the writing process}

During the preparation of this work the authors used Claude Sonnet 4 for grammar checks and slight reformulations. After using this tool, the authors reviewed and edited the content as needed and take full responsibility for the content of the published article.

\printcredits

\bibliographystyle{model1-num-names}

\bibliography{MPCforSFC}


\end{document}